\documentclass[a4paper,11pt]{article}
\usepackage{jinstpub} 

\usepackage{graphicx}
\usepackage{hyperref}
\usepackage{amsmath}
\usepackage{color}
\usepackage[caption=false]{subfig}

\title{\boldmath Spot-Based Measurement of the Brighter-Fatter Effect on a Roman Space Telescope H4RG Detector and Comparison with Flat-Field Data.}

\author[a,b,c,1]{Andrés A. Plazas Malagón,\note{Corresponding author.}}
\affiliation[a]{Kavli Institute for Particle Astrophysics and Cosmology, Stanford University, Stanford, CA, USA}
\affiliation[b]{SLAC National Accelerator Laboratory, Menlo Park, CA, USA}
\affiliation[c]{Department of Astrophysical Sciences, Princeton University, Peyton Hall, Princeton, NJ, USA}

\author[d]{Charles Shapiro,}
\affiliation[d]{Jet Propulsion Laboratory, California Institute of Technology, Pasadena, CA, USA}

\author[e]{Ami Choi,}
\affiliation[e]{NASA Goddard Space Flight Center, Greenbelt, MD, USA}

\author[f,g,h]{Chris Hirata}
\affiliation[f]{Center for Cosmology and AstroParticle Physics (CCAPP), The Ohio State University, Columbus, OH, USA}
\affiliation[g]{Department of Physics, The Ohio State University, Columbus, OH, USA}
\affiliation[h]{Department of Astronomy, The Ohio State University, Columbus, 43210, USA}

\emailAdd{plazas@slac.stanford.edu}

\abstract{We present the measurement and characterization of the brighter-fatter effect (BFE) on a NASA Roman Space Telescope development Teledyne H4RG-10 near-infrared detector using laboratory measurements with projected point sources. After correcting for other interpixel non-linearity effects such as classical non-linearity and inter-pixel capacitance, we quantify the magnitude of the BFE by calculating the fractional area change per electron of charge contrast. We also introduce a mathematical framework to compare our results with the BFE measured on similar devices using autocorrelations from flat-field images. We find an agreement of $18 \pm 5\%$ between the two methods. We identify potential sources of discrepancy and discuss future investigations to characterize and address them.}


\begin{document}

\maketitle
\flushbottom

\section{Introduction}

The Nancy Grace Roman Space Telescope \citep{spergel15} (Roman) is a next-generation space observatory, with a 100 times larger field-of-view than the Hubble Space Telescope, designed to address a wide range of astrophysical questions, from the search for exoplanets to the study of the dark universe. The primary instrument of the telescope is the Wide-Field Instrument (WFI) which will provide wide-field imaging and multi-object, slitless spectroscopy over a $0.281$ deg$^2$ field-of-view, and with a focal ratio of f/7.9. The WFI will be equipped with 18 state-of-the-art Teledyne 4k$\times$4k near-infrared (NIR) H4RG-10 detectors\footnote{Complementary metal-oxide-semiconductor (CMOS) arrays, with the  H4RG-10 readout circuit hybridized to the HgCdTe layer.} with a pixel size of 10 $\mu$m, {and a typical pixel full well of $\approx$ 100000 electrons at a bias voltage of 1V}. Roman will conduct a High Latitude Wide Area Survey (HLWAS) \citep{dore18} that aims to constrain the evolution of the dark energy equation of state and variations from the theory of general relativity using high-quality imaging and accurate management of systematic errors. This survey encompasses 4 near-infrared bands for photometry, has a pixel scale of $0.11$ arcseconds, and includes grism spectroscopy with a resolution of approximately 600. 

Unlike charge-coupled devices (CCDs), these detectors offer a non-destructive readout capability, enabling the sampling of the detector at {equidistant} points in time. {In HXRG detectors, charges generated by light are collected in the depletion region at the p-n junction within the detector layer, causing a voltage change that is measured using non-destructive sampling techniques. This detector layer is linked to a readout integration circuit, which relays the charge to external electronics for conversion into digital form. \citep{rieke07,mosby20}} Importantly, these detectors will be employed for the first time to measure weak lensing of the large-scale structure of the universe, providing a significant advancement in their application within astrophysical research.

The nonlinearity effects that arise in these detectors, namely classical nonlinearity (CNL, \cite{plazas16}), interpixel capacitance (IPC, \cite{kannawadi15}), and the brighter-fatter effect (BFE, \citet{plazas17,plazas18,hirata20,choi20,freudenburg20,givans21}), are all explored as interpixel nonlinearity (IPNL) {in} \citet{givans21}. Understanding and mitigating these effects are crucial because the goal of the HLWAS of the Roman mission is to optimize the results of cosmic shear analysis, where small shear signals necessitate strict control over systematic shape errors at a level of approximately 10$^{-4}$. IPNL has a direct impact {on any light distribution on the detector that is not spatially homogeneous and, in particular,} on the measurements of bright stars, influencing the point spread function (PSF) and the subsequent correction of galaxy shapes. Failure to address these errors properly can lead to propagated inaccuracies throughout the analysis pipeline, affecting the amplitude and clustering of dark matter power spectrum parameters {\citep{choi20,hirata20,mandelbaum15,des22}}. Therefore, a comprehensive investigation of IPNL and, in particular, the BFE in the H4RG-10 detectors is essential to ensure accurate and precise weak lensing measurements in the Roman mission.

\begin{figure}
\includegraphics[width=4.5in]{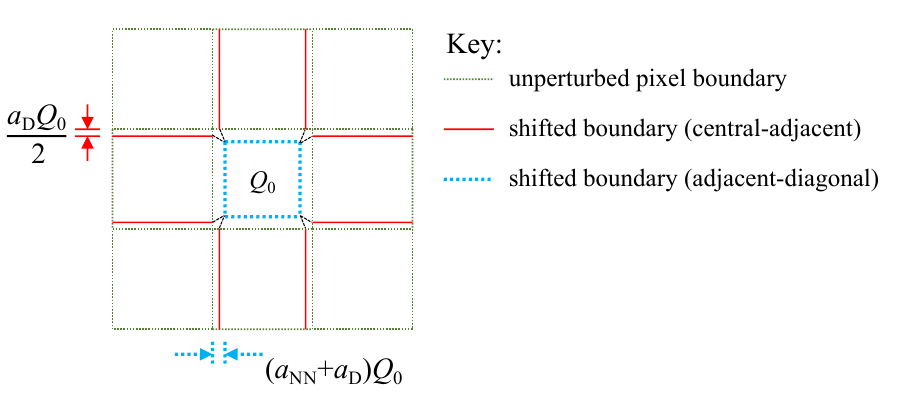}
\caption{\label{fig:shift}The pixel boundary shifts for a ${\cal C}_{4v}$-symmetric BFE kernel. These shifts are used to estimate the relation of the Antilogus coefficients $a_{\Delta i,\Delta j}$ to $B$ in Appendix~\ref{app:compare}.}
\end{figure}

{The brighter-fatter effect is a phenomenon identified first in CCDs that causes images of point sources (such as stars) to appear larger as charge accumulates.
\citet{plazas17} (P2017 hereafter) propose a mechanism for the manifestation of the BFE in HXRG detectors: as charge accumulates in a pixel, it alters the substrate voltage, causing the pixel's depletion region to contract. This contraction could be more pronounced relative to an adjacent pixel if there's a localized concentration of light, potentially causing newly generated charge near the boundary of two pixels to be more likely collected by the pixel with the larger depletion region. Consequently, the collection zones of the pixels physically shift based on the contrast in signal between neighboring pixels. This results in point source images having a fluence-dependent charge redistribution, making brighter sources appear larger (see Figure \ref{fig:shift}, Figure 4 of \citet{plazas17}, and Figure 2 of \citet{mosby20}).}

P2017 {also} characterized the BFE in the H1R detector of the Hubble Space Telescope Wide Field Camera 3 infrared channel using stars from the Omega-cen globular cluster, while \citet{plazas18} (P2018 hereafter) studied this effect using projected point sources on an H2RG-18 detector (18 $\mu$m pixel size) in the controlled environment of a laboratory set up.  As a complement to these studies, \citet{hirata20,choi20,freudenburg20,givans21} study the IPNL and the BFE in a different H4RG-10 detector from this study using flat-field images (with uniform illumination) spatial and temporal autocorrelations, providing a robust mathematical framework to characterize this effect.  

In this paper, we present measurements of the BFE in a 32-channel H4RG-10 Roman Space Telescope development detector\footnote{SCA 18241} using projected spots from data taken at NASA's Jet Propulsion Laboratory's Precision Projector Laboratory (PPL, \citet{shapiro2018}), in a similar study to the work performed in P2018 on an H2RG-18 detector. Section 2 describes the data taken at the PPL, while Section 3 reviews our data analysis methods, including corrections for linear inter-pixel capacitance and classical non-linearity, as well as the metrics defined in P2018 to quantify the magnitude of the BFE using spot data. Section 4 presents our results and comparison to measurements of the BFE with flats correlations performed in similar H4RG-10 detectors by \citet{hirata20,choi20,freudenburg20,givans21}, using the mathematical framework introduced and developed in Appendix~\ref{app:compare}. We conclude in section 5 with a discussion of possible sources of discrepancies between the two methods and future work.

\section{Data}

 The PPL is a detector characterization facility that addresses detector-related risks to ongoing or proposed missions through emulation experiments. The PPL testbed is a one-to-one imager that focuses customized astronomical scenes onto detectors from wavelengths ranging from $0.3$ to $2$ $\mu$m and at focal ratios of f/8 or slower. The PPL can focus over 10,000 images per scene and can rapidly map photometric, astrometric, and PSF shape variations over an entire detector. The PPL has been used to investigate the precision achievable for spectrophotometry of transiting exoplanets, the impact of sub-pixel response variations on photometry for the Euclid mission, and the efficacy of an optimal recombination algorithm for under-sampled images. The PPL is integrated thanks to a multi-disciplinary collaboration of astronomers, detector experts, and optical engineers at JPL and Caltech Optical Observatories.

\begin{figure}
\centering
    \includegraphics[width=0.48\textwidth]{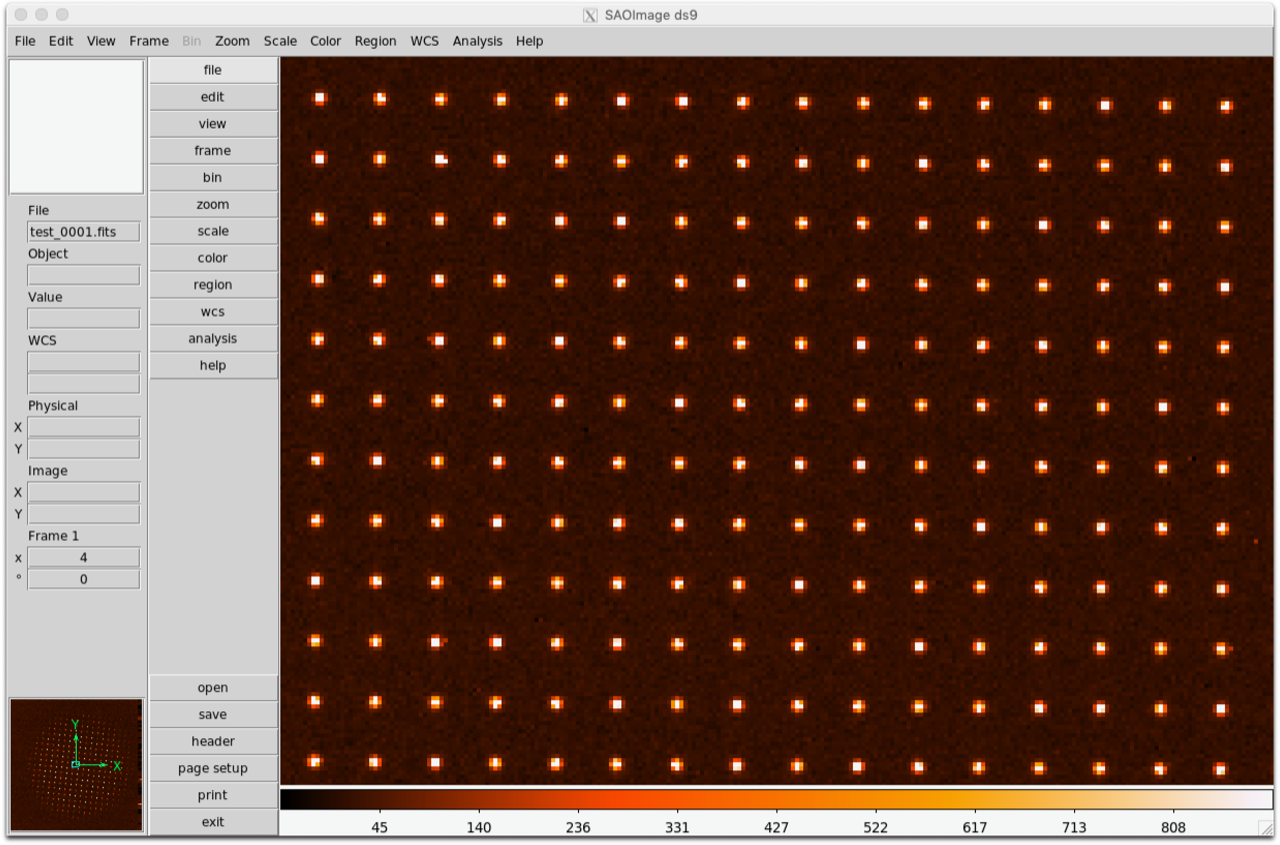}
    \caption{ Approximately 17000 spots projected on H4RG-10 detector used in this study at the Precision Projector Laboratory, with a spot pitch of approximately 146 $\mu$m, taken in the Y band (Roman WFI F106 band.}
    \label{fig:ppl_spots}
\end{figure}

We used a setup that allowed us to illuminate a 32-channel H4RG-10 detector with approximately 17000 point sources of various brightness levels, as well as acquire dark and flat calibration images. The detector used has reference pixels along the outer four rows and columns, however, it should be noted that the reference pixels were not subtracted due to noise: we observed significant 1/f noise that can cause the border rows to not track the rest of the detector. Future work may include exploring the use of techniques to mitigate the impact of this type of noise, such as the ones presented in \citet{rauscher22}. 

We took 50 sample-up-the-ramp images of dark, flat-field (uniform illumination), and spot images (point sources) at f/8 in the Y-band (Roman F106 band\footnote{\url{https://roman.gsfc.nasa.gov/science/WFI_technical.html}}), with the exposure time of the dark images matching that of the flat-field images. {The typical pixel full-well is about 100k e$^-$ \citep{mosby20}, and the maximum value of the median of the spots used in the BFE analysis in this work is about 75\% of this typical full well value. When the pupil is set at f/8 and illuminated at 1 $\mu$m, the optical Point Spread Function (PSF) measures 8µm at full width at half maximum (fwhm). However, when factoring in laboratory seeing conditions and charge diffusion, the overall PSF increases to approximately 10µm. Given that the pixel size is also 10 $\mu$m, the spots are undersampled, which restricts us to spatial information at scales of about 5 $\mu$m.} 

We selected an exposure time and lamp intensity that provided a reasonable number of frames per exposure (9 frames, excluding the reset frame) {and spots and flats reaching approximately 75\% and 55\% of the typical pixel full well}. We discarded the zeroth frame to {mitigate} structures from the channel boundaries and the reset transients. These exposures provided data for the brighter-fatter analyses in this study and potential sub-pixel variation mapping analyses (\emph{e.g.,}, \citet{shapiro_intra18}). During data taking, we observed peak-to-peak temperature variations of less than 8 mK. The Y-band measured background in the central detector region, where the background is lowest, was of approximately 12 Analog-to-Digital Units per second (ADU/s) or 26.4 e$^-$/s (for a mean detector gain of $2.2$ e$^-$/ADU). Additionally, we used a photodiode to monitor the light source and verify the long-term lamp stability. \footnote{{Previous characterization of the testbed instrumentation has demonstrated the stability of the lamp flux to be better than 0.1\% when regulated via a closed feedback loop with the reference photodiode. Averaging across numerous experimental exposures serves to further reduce any potential impacts stemming from minor fluctuations in the lamp output throughout data collection.}}

\section{Methods}

In this analysis, we mainly follow the methodology in P2017 and P2018. We examine the $3\times3$ pixel region\footnote{{We used the pipelines from P2018, which use a $3\times3$ stamp size. In addition, \citet{choi20} find a rapid fall-off of the BFE beyond the closest 4 pixels ($\sim r^{-5.6\pm0.2}$, for a distance r from the central pixel)}.} or postage stamp at the center of each spot to investigate the BFE in the PPL data, after accounting for known effects such as classical non-linearity (CNL) and inter-pixel capacitance (IPC). We then analyze the spot images for any variations in flux due to charge contrast between pixels. We visually inspect the dark, flat, and spot exposure ramps and discard any outliers, as described in Appendix \ref{ap:cnl}. We then compute the median ramp for each set to improve the signal-to-noise ratio (SNR). The ramp stacking process also helps mitigate lamp fluctuations, image motion, and seeing-related {and electronic read-out} systematic errors. To identify the spots in our images, we utilize { \tt {SExtractor}} \citep{bertin96} on the {last frame} of the median spot ramps. We apply a mask to flag bad pixels, generated from laboratory measurements. This allows us to locate the approximate positions of the spots and calculate the unweighted centroid based on the inner $3\times3$ postage stamp of each spot. Initially, we select sources with centroids within $\pm 0.1$ pixels of the pixel center to maximize the signal contrast and increase the amplitude of the BFE between the central pixel and its nearest neighbors since the amplitude of the BFE depends on the charge contrast between neighboring pixels: in the model of P2017, the shift of the pixel boundary in NIR detectors occurs when the size of the p-n junction depletion region changes due to the presence of charge induced by illumination or the {bias voltage after reset}. In the absence of signal contrast, the pixel boundary is positioned at the midpoint between the edges of the depletion region. However, when one pixel receives a stronger signal compared to the other {and as charges accumulate,  the geometric configurations of the depletion region and its electric fields evolve,} causing the effective boundary between the pixels to move towards the pixel with the higher signal {and thereby inducing the BFE (see Figure 4 of \citet{plazas17} and Figure 2 of \citet{mosby20}).}

\subsection{Linear IPC}

Originating from the capacitance coupling between adjacent pixels in the source-follower, inter-pixel capacitance manifests as a source of crosstalk between neighboring pixels. The signal at a specific pixel is affected by this capacitance coupling \citep{kannawadi15}.  The resulting inter-pixel capacitance effect is represented as a linear effect, described as a convolution of a $3\times3$ kernel with the image, ultimately causing images to appear blurred. We assume that IPC is linear as well as spatially and temporally uniform\footnote{{Although see \citet{choi20} for a map of spatial dependence of IPC at the $\approx$ 0.3 \% level in a similar detector.}}, and we
correct for it by applying the following kernel, estimated by comparing the signal in hot
pixels to that of their nearest neighbors. To correct for the effects of constant inter-pixel capacitance, we used a kernel derived from averaging the signals around approximately 1300 hot pixels: 

\begin{gather}
    K = 
\begin{bmatrix}
0.00185 &  0.01413 & 0.00184 \\
0.01924 & 0.9322 & 0.01831 \\
0.0014  & 0.00948 & 0.00154 \\
\end{bmatrix}
\end{gather}

We deconvolve the IPC kernel at each frame for the darks, flats, and spots stacked ramps. The IPC kernel orientation is such that the rows and columns of the matrix correspond to the rows and columns of the images as in e.g. Figs. \ref{fig:ppl_spots} and \ref{fig:bad_vs_good_NL_coeffs}.

\subsection{Classical non-linearity}

To correct for classical non-linearity (see, e.g., \citet{plazas16}), P2017 and P2018 use the analytic solution to a quadratic fit that describes the measured signal {with no signal contrast} per pixel in electrons\footnote{We convert the ADU numbers from raw images to electrons by using an IPC-corrected
average mean gain of 2.2 e-/ADU.} as a function of time $t$:
\begin{equation}
    S = C_0 + C_1t + C_2(C_1t)^2
    \label{eq:quadratic}
\end{equation}
In equation \ref{eq:quadratic}, $C_0$ represents an offset in the ramp due to a reset voltage, and $C_1t \equiv Q_{\textrm{L}}(t)$ is the linear flux component of the signal. P2018 derived the $C_2$ quadratic coefficient from the flat-field ramps (to be used in the CNL correction for the dark, spot, and flat images themselves), but the constant and linear parts of Eq. \ref{eq:quadratic} (the  $C_0$ and $C_1$ coefficients) for the spot and dark images were derived from {quadratic} fits to their own ramps.  In this work, we follow a different approach to perform the classical non-linearity correction, as shown in \citet{freudenburg20,troxel23}. We first take the median of  the measured {flat-field} ramps, rejecting outlier ramps based on how each one compared to {mean ramp}, and then, following \citet{freudenburg20}, we fit an unweighted polynomial to each pixel of the median stacked ramp:

\begin{equation}
S_{\rm{median}} = \sum_{i=0}^p k_j t^j, 
\end{equation}

Here, $p$ represents the polynomial order, chosen to be $p=3$ in this work. We then normalize the slope (linear part) of the polynomial to one by defining $c_j \equiv k_j / k_1^j$ (as in \citet{freudenburg20}). The non-linearity coefficients, $\beta_j$ are therefore given in terms of the average gain $g$ (e$^-$/ADU), and the normalized polynomial coefficients, $c_j$, by: 

\begin{equation}
    \beta_j = -\frac{c_j}{g^{j-1}}
    \label{eq:NL_corr}
\end{equation}

We used the implementation of this non-linearity correction from the Instrument Signature Removal and Calibration-Products Production code of the Legacy Survey of Space and Time (LSST) Science Pipelines\footnote{\url{https://github.com/lsst/cp_pipe},\ \url{https://github.com/lsst/ip_isr}, {\tt{w$\_$2023$\_$38}}}. {The left panel of Fig.~\ref{fig:flat_outliers_and_cnl_res} in Appendix \ref{ap:cnl} shows how the polynomial correction used in this work mitigates the impact of CNL on the $f_N$ metric used to quantify the BFE (this metric is defined in the next section). However, future analyses should explore the efficacy of other potential CNL correction algorithms (e.g., the methods proposed in \citet{canipe17} and \citet{rauscher19}).}

\subsection{BFE Metrics}
\label{sec:3.3}

We subtract the CNL-corrected dark ramps from the CNL-corrected spot ramps, and we convert the signal in each pixel from electrons (e$^-$) to a flux in electrons per second (e$^-$/s). As in P2018, we define the relative flux deviation in a particular pixel as follows:

\begin{equation}
f_{\rm N}(i) = \frac{F_{i} - F_1}{F_{\star}}
\label{eq:fn}
\end{equation}

with a normalization factor (the time-averaged factor of the 9 pixels in the postage stamp):

\begin{equation}
F_{\star}= \frac{\sum_{p=1}^9 \left( S_{p,M-1} - S_{p,1}\right)}{\Delta t (M-2)}
\label{eq:fstar}
\end{equation}

In Equations \ref{eq:fn} and \ref{eq:fstar}, $F_i$ is the pixel flux obtained from taking the difference between frames $i+1$ and $i$ in the data ramp, $S_{p,i}$ is the signal in the $i$-th frame of the $p$-th pixel, $M = 9$ is the initial total number
of frames, and $\Delta t$= 2.73 s is the frame time. 
For an ideal response, the relative flux deviation  $f_{\rm N}(i)$ should be consistent with zero. 

Following \citep{plazas18}, we estimate the change in pixel area by modeling the current in the central bright pixel as:

\begin{equation}
\frac{dQ}{dt} = F(t) = F_0 \left(1+ BQ_{\rm c}(t) \right) =  F_1 \left(1+ BF_{\rm c}t) \right)
\label{eq:current}
\end{equation}

In Equation \ref{eq:current}, $F_1$ represents the unchanged flux in the reference frame, and $Q_{\rm c}$ and $F_{\rm c}$  are the
mean charge and flux contrast between the central pixel and its four nearest neighbors,
respectively. The coefficient $B$ models the fractional area change per electron of contrast, assuming that the flux across the central pixel is spatially constant (square top-hat profile).  This overestimates the charge redistribution (and thus underestimates $B$) relative to a model that accounts for the shape of the PSF at the pixel boundaries.  Combining Equation \ref{eq:fn} and \ref{eq:current}, we have: 

\begin{equation}
    f_{N} = \frac{B F_1 F_{\rm c} \Delta t}{F_{\star}} i
    \label{eq:b}
\end{equation}

\section{Results}

\subsection{Measurement of B coefficient}
\label{sec:meas_b}
\begin{figure}
    \begin{minipage}{\textwidth}
        \centering
        \includegraphics[width=0.8\textwidth, page=1]{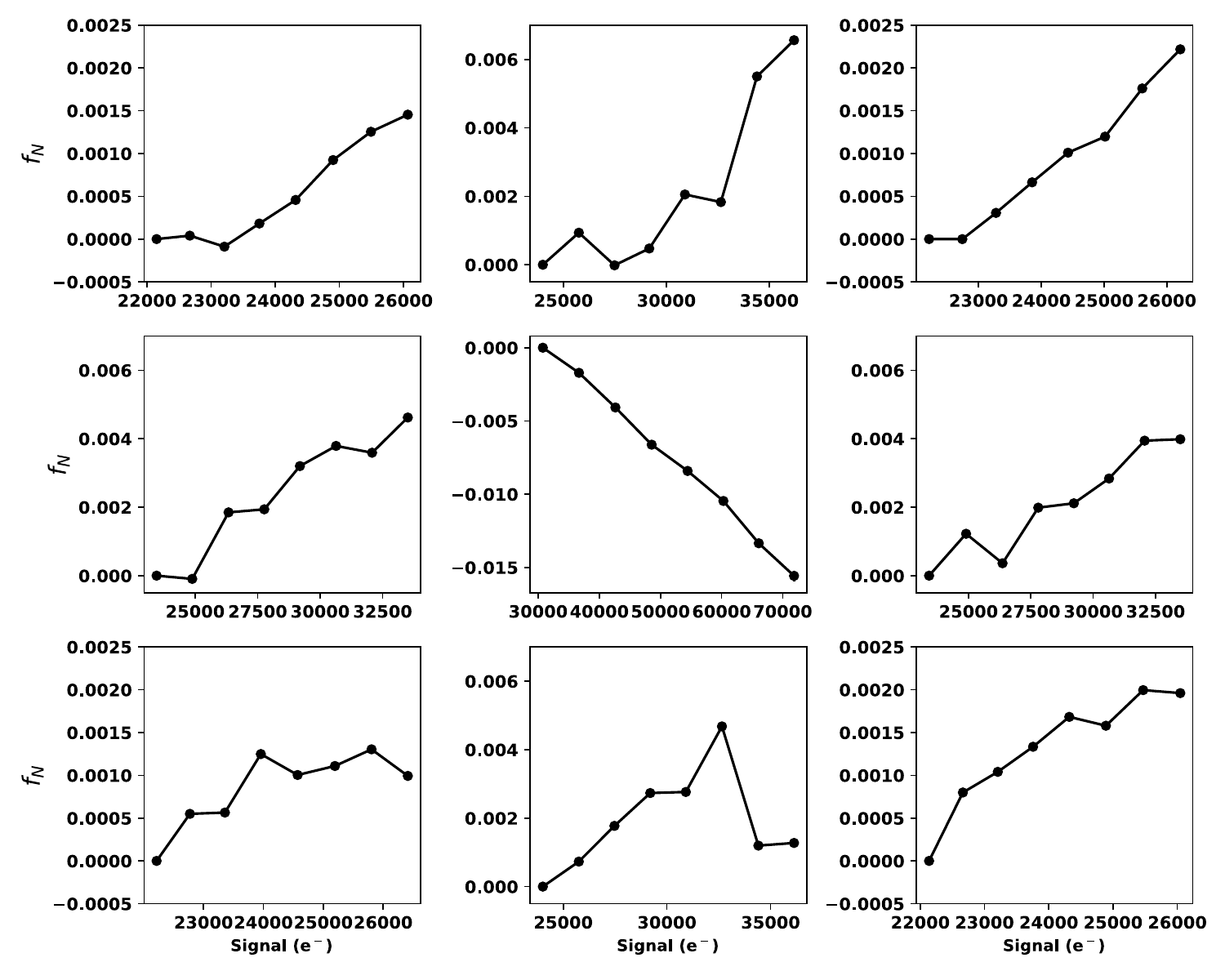}
    \end{minipage}
    
    \vspace{1em} 
    
    \begin{minipage}{0.48\textwidth}
        \centering
        \includegraphics[width=\linewidth, page=3]{fn_combined_2023SEP23.pdf}
    \end{minipage}
    \hfill
    \begin{minipage}{0.48\textwidth}
        \centering
        \includegraphics[width=\linewidth, page=4]{fn_combined_2023SEP23.pdf} 
    \end{minipage}
    
    \caption{\emph{Top}:  Average normalized change in flux relative to the start of the exposure, $f_N$ for each pixel in a 3-pixel by 3-pixel region. The data is averaged over 50 exposures and 2957 spots with a centroid $< 0.1$ pixels away from a pixel center. The error bars (standard error of the mean) are smaller than the symbols. \emph{Bottom left}: Histogram of the $B$ coefficient (see section \ref{sec:3.3} in the text) after 3-sigma clipping. \emph{Bottom right}: Charge re-distribution from the central pixel (red stars) of the plot in the top panel into the surrounding 8 pixels (blue-filled circles.)}
    \label{fig:three_figures}
\end{figure}

We use the CNL and constant IPC corrected median spot ramp to calculate the $B$ coefficient from P2018 and Equation \ref{eq:b}, which is interpreted as the area change $dA/A$ per e$^{-}$ of charge contrast between the central pixel and its neighbors for the median spot of all those spots in the detector that have a {centroid within 0.1 pixels of a pixel (maximum deviation)}. The magnitude of the BFE in HXRG detectors as manifested by the change of effective pixel boundaries is proportional to the charge contrast between pixels, as proposed by the model in P2017 and P2018. The mean magnitude of the $B$ coefficient, from measuring $f_N$ over 2957 spots as in Equation \ref{eq:b}, is found to be $-1.038 \pm 0.0260$ ppm/e$^{-}$ (bottom left panel of Fig.~\ref{fig:three_figures}) for the H4RG-10 detector with a pixel size of 10 $\mu$m used in this study. P2018 find a less negative coefficient in an H2RG-18 detector with a larger pixel size of 18 $\mu$m  ($B = 0.41 \pm 0.0076$ ppm/e$^{-}$), although with a smaller scatter. Fig.~\ref{fig:three_figures} (top and bottom right panels) also illustrates the charge-conservation plots shown in Figs. 3 and 4 of P2018, where the relative charge deficit in the central pixel and the corresponding redistributed excess in the surrounding pixels is shown using the same $f_N$ metric as in this work, quantifying the relative flux in the frames of an image ramp relative to the first usable frame.   

\subsection{Comparison with BFE characterization from flat-field images}

\citet{hirata20,choi20,freudenburg20,givans21} performed a general IPNL analysis (including the BFE) in candidate Roman H4RG-10 detectors (different from the one used at the PPL for this study) using pixel spatial autocorrelations in flat-field images, as well as correlations between the different pixel values as the charge integrates in time up the sampling ramp. They characterize the BFE in terms of the phenomenological model originally proposed by \citet{antilogus14,guyonnet15} in fully-depleted, thick CCDs \citep{Downing_2006,antilogus14,guyonnet15,gruen2015,Astier_2019,Astier_2023}. In this model, the change in the effective area of a particular pixel depends on the charge in its neighboring pixels as follows \citep{hirata20}:

\begin{equation}
\mathcal{A}_{i, j}=\mathcal{A}_{i, j}^0\left[1+\sum_{\Delta i, \Delta j} a_{\Delta i, \Delta j} Q(i+\Delta i, j+\Delta j)\right]
\label{eq:antilogus}
\end{equation}

In Eq.~\ref{eq:antilogus}, $Q(i, j)$ is the charge in pixel $(i, j)$, and $a_{\Delta i, \Delta j}$ denotes the proportionality coefficients. The $B$ coefficient derived in the previous section and in the P2017 and P2018 analyses from focused light (stars and projected spots) can be related to the $a_{\Delta i, \Delta j}$ from the Antilogus model, derived from flat-field images with uniform illumination. Under the assumption of a symmetrical BFE kernel of the form

\begin{equation}
\left[\begin{array}{ccc}
a_{\rm D} & a_{\rm NN} & a_{\rm D} \\
a_{\rm NN} & -4a_{\rm D}-4a_{\rm NN} & a_{\rm NN} \\
a_{\rm D} & a_{\rm NN} & a_{\rm D}
\end{array} \right],
\end{equation}
Appendix~\ref{app:compare} demonstrates that, for a Gaussian spot of RMS size per axis $\sigma$, the $B$ coefficient is given by 
\begin{equation}
B = -\frac{2\sqrt{2}\,e^{-1/(8\sigma^2)}}{\sqrt\pi\,\sigma\,{\rm erf}\, \frac1{2\sqrt2\,\sigma}}
\left[a_{\rm NN}
+ \frac12 \left( 1 + 
\frac{
 {\rm erf}\, \frac3{2\sqrt2\,\sigma}
}{ {\rm erf}\, \frac1{2\sqrt2\,\sigma}}\right)
a_{\rm D}\right];
\label{eq:b_vs_a}
\end{equation}
for an Airy spot the coefficients of $a_{\rm NN}$ and $a_{\rm D}$ can be obtained numerically.

The profile of the spots projected at the PPL on the H4RG detector used in this work can be modeled with an Airy profile, which can be approximated as Gaussian with $\sigma = 0.42 \lambda F$, where $\lambda$ and $F$ denote the illuminating wavelength and the f-number of the optical setup, respectively. Using values of $a_{\rm NN} = 0.29$ ppm, $a_{\rm D} = 0.038$ (Fig. 5 of \citet{choi20}, for SCA 18237), $\lambda = 1.060\ \mu$m (central wavelength of Roman WFI F106 band\footnote{\url{https://roman.gsfc.nasa.gov/science/WFI_technical.html}}), and $F=8$ \citep{shapiro2018}, we obtain a value of $B = -0.66$ ppm/e$^{-}$ from Eq.~\ref{eq:b_vs_a} {as a first estimation for the value of this coefficient. However,} this is an underestimate of the magnitude if there are additional contributions to the spot size from, e.g., {charge diffusion}. These additional contributions were not measured for {the detector from \citet{choi20}}, SCA 18237, but they were measured for the flight SCAs during acceptance testing. We can thus {produce an updated prediction of} $B$ using the equations from Appendix~\ref{app:compare} for an Airy disc convolved with a Gaussian charge diffusion, and using the BFE and charge diffusion parameters measured for the 18 flight SCAs during the acceptance testing ({as presented in \citet{troxel23,givans21,mosby20}}: see \citet{mosby20} for an overview of the flight detector program; \citet{givans21} for the charge diffusion measurement technique using short-wavelength flat field correlations; and \citet{troxel23} for a description of the SCA summary files). The result is $B=-0.85\pm0.05$ ppm/e$^{-}$, where the error bar indicates the error on the mean from the dispersion of the individual SCA results.
This prediction is $18\pm 5$\% different from the mean value measured in this study ($B=-1.038 \pm 0.0260$ ppm/e$^{-}$, section~\ref{sec:meas_b}). 

\section{Conclusion}

In this study, we conducted measurements of the brighter-fatter effect (BFE) in an H4RG-10 detector using projected spots in JPL's Precision Projector Laboratory (PPL) at Caltech. Our methodology follows the approach used by \citet{plazas18} (P2018) in their investigation of a similar H2RG-18 detector at the PPL. We obtained a $B$ coefficient of $1.038 \pm 0.0260$ ppm/e$^{-}$, which quantifies the change in photo-generated current in the central pixel of a spot per electron contrast between that central pixel and its neighboring pixels.

Comparing our results to P2018, who examined a detector with a larger pixel size, we find a more negative $B$ coefficient, {as photo-carriers have a smaller probability of diffusing to neighboring pixels.} However, our measurement errors are larger than those reported by P2018. Further investigations are needed to identify the sources of error and potential systematic uncertainties.

To evaluate the BFE, we compared our spot-based approach (measuring $B$) with the flat field autocorrelation method employed by \citet{hirata20,choi20,freudenburg20,givans21}. The latter method determines changes in the effective area of a pixel as it collects electrons, as described by the model proposed by \citet{antilogus14} and \citet{guyonnet15}. Our analysis revealed an agreement within $18\pm5\%$ between the two methods. 
Potential sources for the remaining discrepancy include: (i) deviations from the Antilogus boundary-shift model; (ii) other forms of non-linear crosstalk (e.g., the vertical trailing pixel effect in the readout integrated circuit [ROIC] \cite{freudenburg20}, which also makes a spot appear larger as the signal level is increased); (iii) chip-to-chip variations in the BFE (different detectors used in this study and the work on flat autocorrelations); (iv) operational conditions, especially for effects that arise in the ROIC; (v) wavelength dependence ({i.e., where the photo-carrier is produced in the detector}); and (vi) image motion contributions in the spot projector.
However, it is worth noting that besides differences in data sets and analysis pipelines, the change in area computed for flat illumination may not align with the weighted $B$ coefficient since the response to focused light can exhibit variations.  Future investigations should incorporate additional null tests to examine potential error sources and further compare the two methods. Additionally, exploring the application of the autocorrelation analysis to spot images could be beneficial. While flat images offer a larger number of pixels for averaging, spot data induces a more pronounced BFE due to its higher contrast, thereby enabling the exploration of contrast-dependent non-linearities.

{The BFE primarily impacts weak lensing measurements (galaxy shape measurements and shear inference) through its effect on the size and ellipticity of the Point Spread Function (PSF), as modeled from bright stars. An accurate model of the PSF is crucial, as it must be effectively deconvolved in the images of faint galaxies during the process of inferring shear.
Current state-of-the-art algorithms for BFE correction in weak lensing galaxy surveys, as delineated in \citet{antilogus14,gruen2015,guyonnet15,coulton18,broughton23}, are employed in ground-based experiments with CCD cameras, such as the Dark Energy Survey (DES), the Hyper Suprime-Cam Survey (HSC), and are considered to be employed in future Stage IV surveys such as the Legacy Survey of Space and Time (LSST) and the High Latitude Wide Area Survey by Roman. These methodologies entail deriving a correction kernel from flat field data and applying it to the point sources from which the PSF model is constructed. This approach assumes that a BFE correction derived from uniform illumination is equally applicable to focused light. 
For space-based measurements, current work (see, \emph{e.g.}, \citet{hirata23}) is in progress to demonstrate that shear measurement algorithms, such as {\tt {Metadetection}} \citep{sheldon20,sheldon23}, can achieve the requisite accuracy with undersampled PSFs from space (using software such as IMCOM \citep{rowe11,yamamoto23} to accurately reconstruct the PSF). 
The present study contributes to a deeper comprehension of the BFE in H4RG detectors that will be utilized for weak-lensing science with the Roman Space Telescope. Given the magnitude of the BFE and the observed level of discrepancy (approximately $18\pm5\%$), it may be necessary to continue with the point source measurement method, potentially including on-sky observations, as well as flat field corrections, to refine the BFE correction for cosmic shear science. However, future work should also thoroughly investigate the potential discrepancy sources suggested above to resolve any measurement or systematic discrepancies.}

\section{Acknowledgments}

The work of AAPM was supported by the U.S. Department of Energy under contract number DE-AC02-76SF00515.
Part of this research was carried out at the Jet Propulsion Laboratory, California Institute of Technology, under a contract with the National Aeronautics and Space Administration (80NM0018D0004). We thank B. Rauscher (NASA Goddard Space Flight Center) and G. Mosby (NASA Goddard Space Flight Center) for useful discussions and for providing details about the Roman in-flight conditions. 

\appendix
\section{Classical non-linearity correction}
\label{ap:cnl}

To calculate the $B$ histogram, we produce a median file for each of the file types (spots, darks, and flats), and we correct for classical non-linearity (CNL) in each one. 
We use the median flat ramp to derive the CNL correction coefficients, up to the third order, in Eq.~\ref{eq:NL_corr}.  Before producing the median flat image, we identify outlier ramps by calculating the mean pixel value per ($4096 \times 4096$) frame for each of the initial 50 ramps, forming { \tt{mean\_ramp\_i}},  for $i \in [1, 50]$, and then calculating the ratio {\tt{mean\_ramp\_i}} / {{\tt meanAllRamps}} - 1,  where {{\tt meanAllRamps}}  is the mean of all the individual signal ramps. 
\begin{figure}
\centering
    \includegraphics[width=0.45\textwidth, page=1]{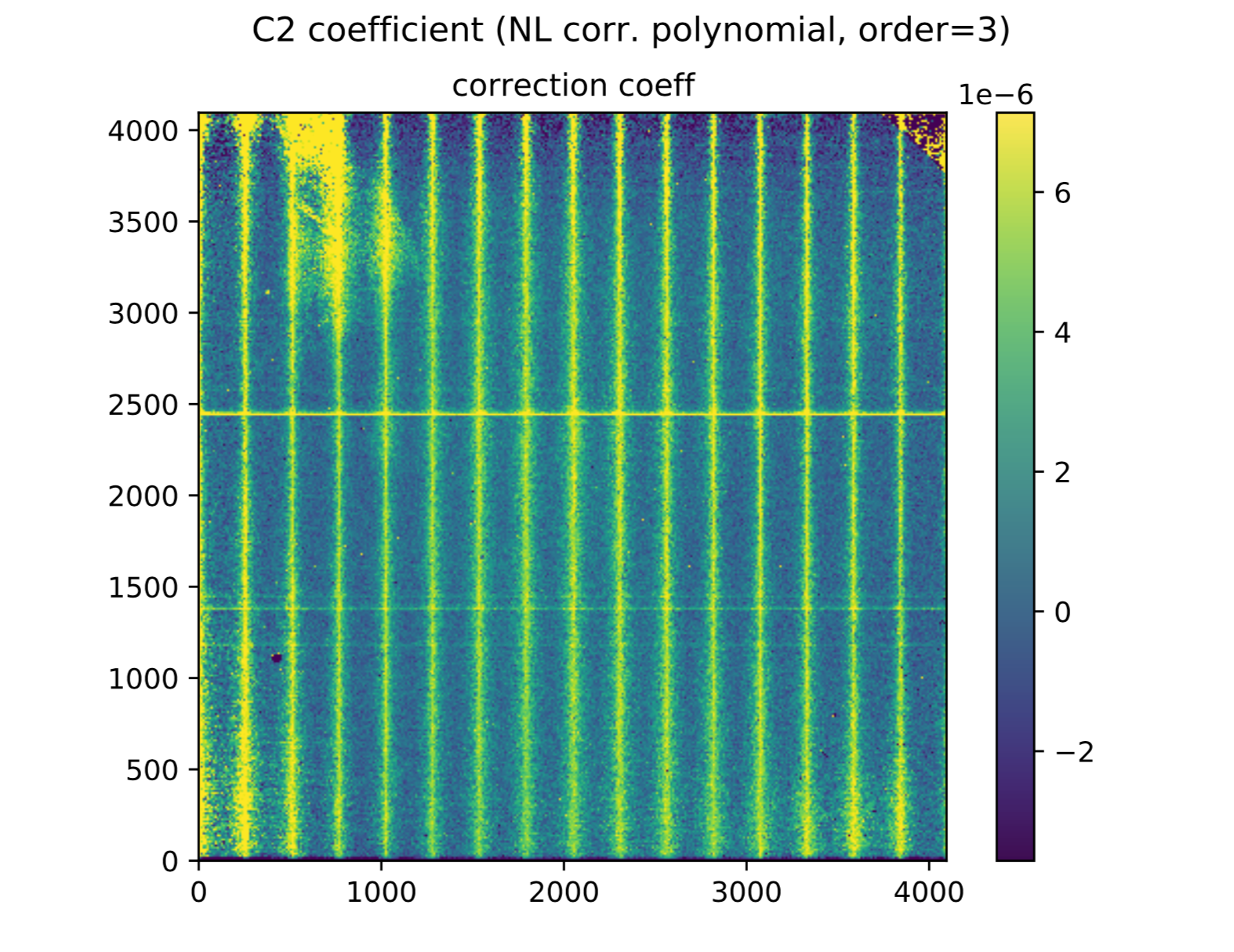}
    \includegraphics[width=0.45\textwidth, page=2]{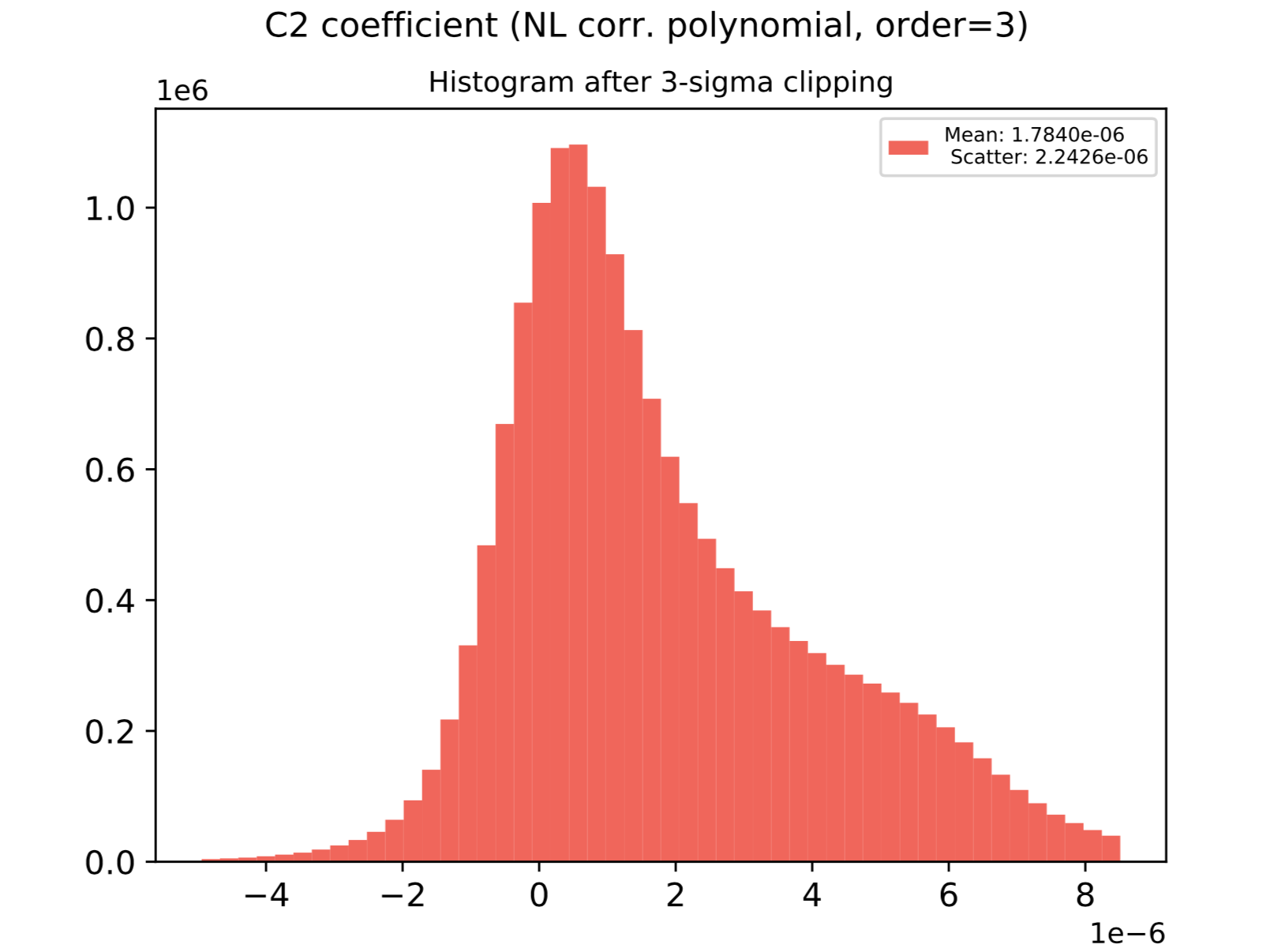}
    \caption{The classical non-linearity correction is affected when the zeroth sample-up-the-ramp frame from the flats is not discarded: channel boundaries and horizontal lines from the reset transient are left imprinted in the CNL correction coefficients maps, and a skewed coefficient distribution is observed. This figure displays the effect on the CNL correction quadratic C2 coefficients; a similar effect is found in the cubic matrix, C3 (not shown).}
    \label{fig:bad_vs_good_NL_coeffs}
\end{figure}
\begin{figure}
\centering
    \includegraphics[scale=0.4]{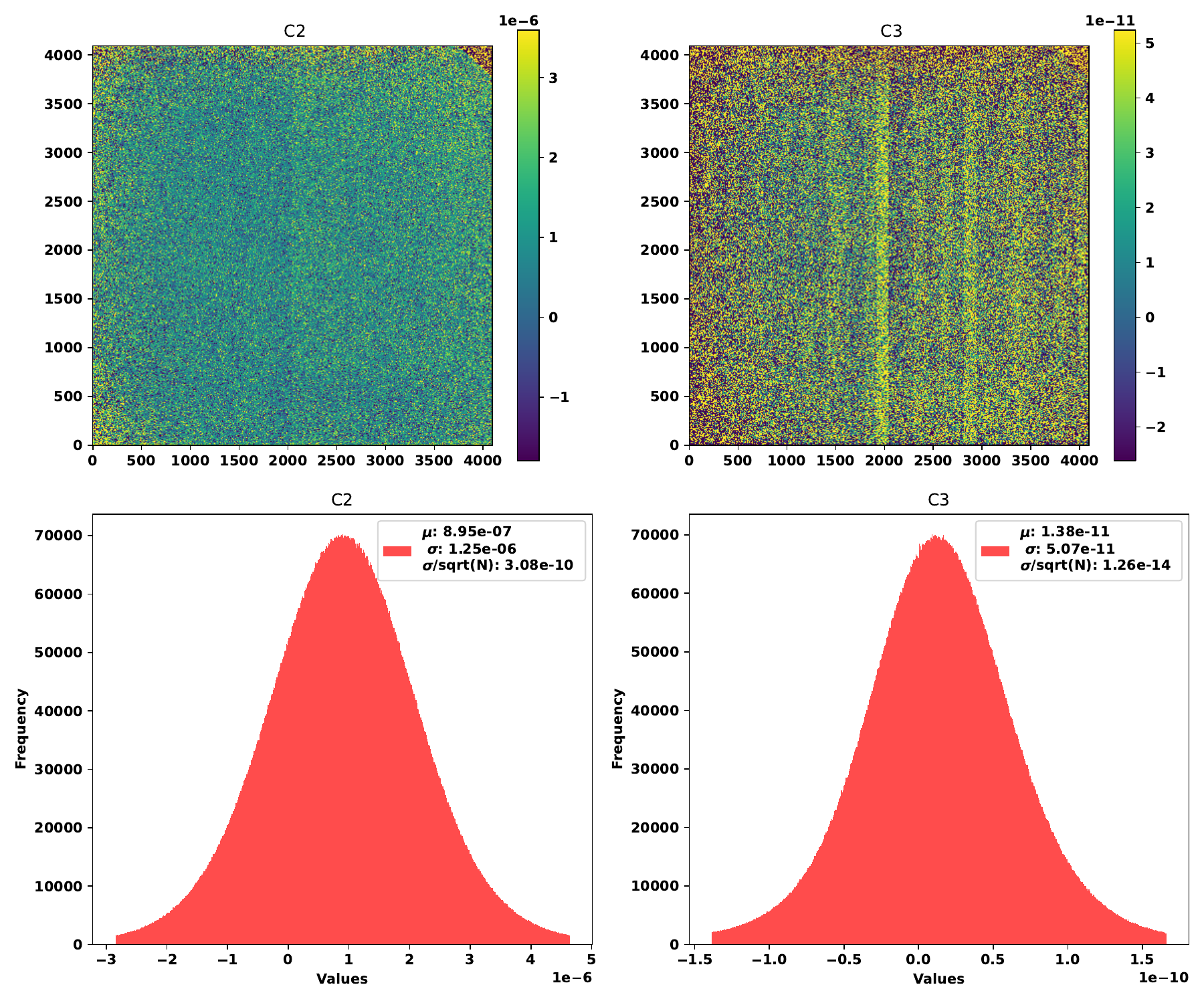}
    \caption{ Classical non-linearity correction coefficients after 3-sigma clipping, from a third-order polynomial fit to each pixel ramp in a median flat image cube, {discarding the zeroth frame}.}
    \label{fig:CNL_coeffs}
\end{figure}
\begin{figure}
\centering
    \includegraphics[width=0.42\textwidth, page=2]{fn_combined_2023SEP23.pdf}
     \includegraphics[width=0.42\textwidth]{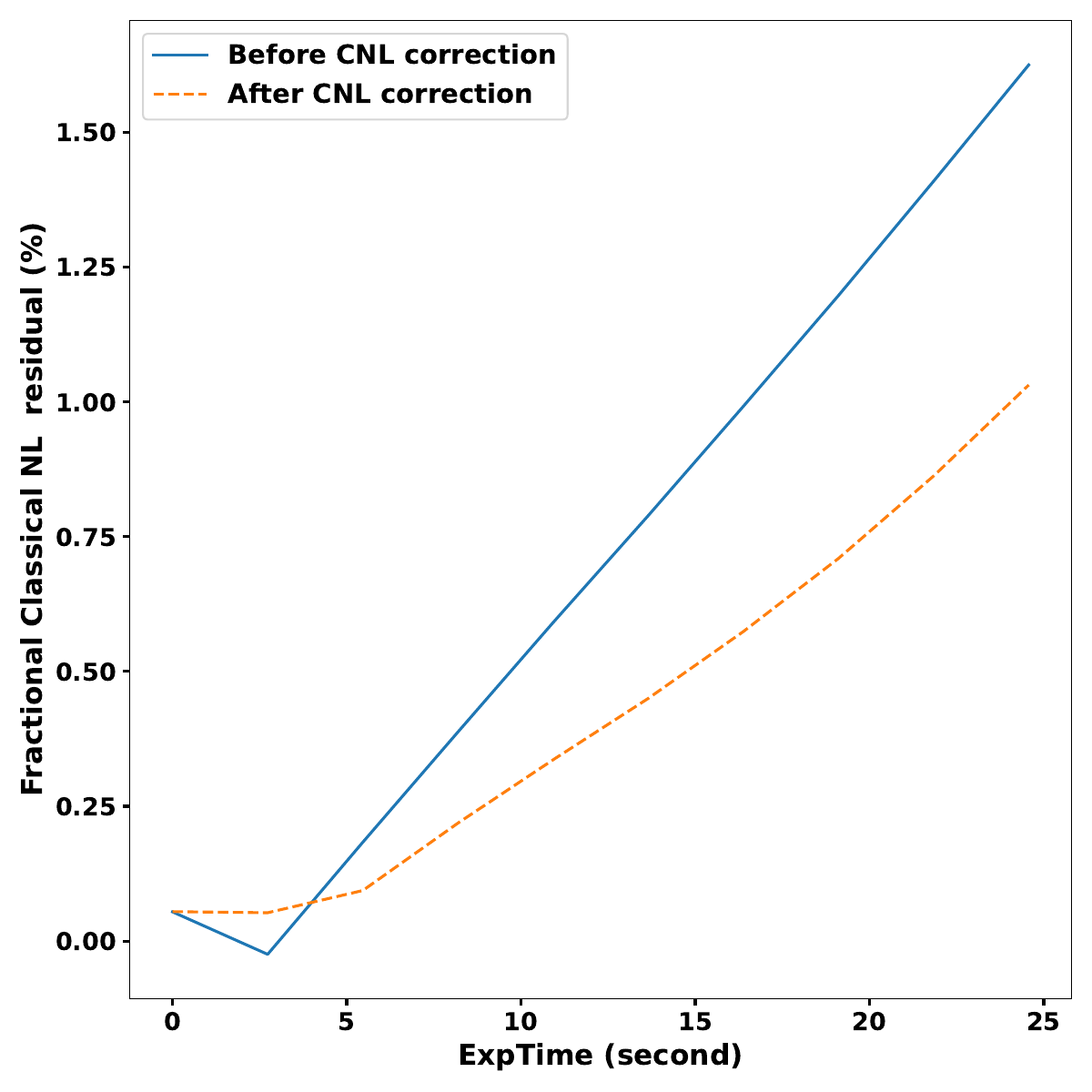}
    \caption{\emph{Left}: Average $f_N$ metric for the central spots and flat-field ramps. The CNL correction ``flattens" the curve, as expected, since the BFE signal should be small due to the low charge contrast between pixels in flats. \emph{Right:} Classical non-linearity residuals, defined as the absolute fractional deviation of the signal from the linear component of the cubic fit to the signal, as a function of time for a median flat ramp.}
    \label{fig:flat_outliers_and_cnl_res}
\end{figure}
This process led us to discard several flat-field ramps\footnote{Ramps number 1, 11, 45, 46, 47, 48, and 49 out of the 50 flat-field ramps} before median-stacking the data cubes (ramps) since they seemed to deviate from the rest.  Each data cube has dimensions ({{\tt N\_FRAMES}}, {{\tt X\_SIZE}}, {{\tt Y\_SIZE}}), which in our case is equal to ($9, 4096, 4096$). The raw data originally consisted of 10 frames in all cases, however, we discarded the zeroth frame from our analyses to avoid effects such as residuals from transients. Fig.~\ref{fig:bad_vs_good_NL_coeffs} {and Fig.~\ref{fig:CNL_coeffs}} illustrate the impact of not discarding this zeroth frame on the matrix of quadratic CNL-correction coefficients (a similar effect can be seen in the cubic coefficients). 

We then use the median flat to calculate the second and third-order (cubic polynomial) CNL matrices and use those coefficients to correct for CNL in the median-stacked flats, darks, and spots cubes which we then use to calculate the $B$ coefficient defined above.  The distributions are illustrated in Fig.~\ref{fig:CNL_coeffs}, and the CNL residuals of the median ramp as a function of the time are shown in the right panel of Fig.~\ref{fig:flat_outliers_and_cnl_res} and are found to be below $1\%$ after correction.  In addition, the left panel of Fig.~\ref{fig:flat_outliers_and_cnl_res} shows the central pixel of the $f_N$ metric for the spots and flat-field ramps. In particular, the metric shows the effect of the CNL correction on flats, for which the BFE $f_N$ metric should tend to zero after classical non-linearity correction as they are exposed to uniform illumination with little charge contrast between neighboring pixels.  

\section{Comparison of flat field and spot-based methods for measuring BFE}
\label{app:compare}

This appendix describes a framework for comparison of two methods for estimating the BFE: the flat field method, and the spot projection method used in this paper. The flat field correlation function measures the Antilogus coefficients $a(\Delta i,\Delta j)$, which describe the change of pixel area when a pixel has collected electrons. The spot projection method measures the coefficient $B$, the change in photocurrent in the central pixel of a spot per electron of contrast between that central pixel and its neighbors.
In all of the following, we use 1 pixel as our unit of length unless otherwise specified.

Let us begin by supposing that $a$ has the form of a ${\cal C}_{4v}$-symmetric\footnote{That is, we impose symmetry including rotations by multiples of 90$^\circ$, as well as reflections across the horizontal, vertical, and 2 diagonal planes. Some constraints are necessary to turn flat field Antilogus coefficients (which describe changes in pixel area) into shifts of pixel boundaries.} $3\times 3$ kernel,
\begin{equation}
a(-1...+1,-1...+1) = \left( \begin{array}{ccc}
a_{\rm D} & a_{\rm NN} & a_{\rm D} \\
a_{\rm NN} & -4a_{\rm D}-4a_{\rm NN} & a_{\rm NN} \\
a_{\rm D} & a_{\rm NN} & a_{\rm D}
\end{array} \right).
\end{equation}
If we follow ${\cal C}_{4v}$ symmetry, then if a pixel contains an electron, then each of its four boundaries moves inward by $a_{\rm NN}+a_{\rm D}$ pixels; and each of the eight boundaries of the adjacent pixels moves by $a_{\rm D}/2$ (see Figure~\ref{fig:shift}).

Now suppose that we illuminate the detector with a spot with intensity $\phi(x,y)$, which is normalized to integrate to 1, {intended to represent both the optical spot, convolved with any charge diffusion}. The fraction of the total flux in a pixel $(i,j)$ at the beginning of the exposure, before the BFE becomes significant, is
\begin{equation}
f_{\rm init}(i,j) = \int_{i-1/2}^{i+1/2} \int_{j-1/2}^{j+1/2} \phi(x,y)\,dx\,dy.
\label{eq:finit}
\end{equation}
After a total signal $Q_{\rm tot}$ (total electrons in all pixels) has been collected, the right boundary ($i+\frac12$) upper limit has moved by an amount
\begin{eqnarray}
\Delta x_{\rm max} &=& Q_{\rm tot} \Bigl\{ (a_{\rm NN}+a_{\rm D})[f_{\rm init}(i+1,j)-f_{\rm init}(i,j)]
\nonumber \\ && ~~~~
+ \frac{a_{\rm D}}2 [f_{\rm init}(i+1,j+1)+f_{\rm init}(i+1,j-1)-f_{\rm init}(i,j+1)-f_{\rm init}(i,j-1)]
\Bigr\},
\label{eq:dxmax}
\end{eqnarray}
and similarly for the other boundaries. The result is a change in the fraction of the flux in a given pixel per total electron collected:
\begin{eqnarray}
\frac{\Delta f(i,j)}{Q_{\rm tot}}
&=& \bigl\{(a_{\rm NN}+a_{\rm D})[f_{\rm init}(i+1,j)-f_{\rm init}(i,j)] + \tfrac12a_{\rm D}[f_{\rm init}(i+1,j+1) + f_{\rm init}(i+1,j-1)
\nonumber \\ && ~~~~
-f_{\rm init}(i+1,j) - f_{\rm init}(i-1,j)]\bigr\} \int_{j-1/2}^{j+1/2} \phi\left( i+\tfrac12,y\right)\,dy
\nonumber \\ &&
+ \bigl\{(a_{\rm NN}+a_{\rm D})[f_{\rm init}(i-1,j)-f_{\rm init}(i,j)]  + \tfrac12a_{\rm D}[f_{\rm init}(i-1,j+1) + f_{\rm init}(i-1,j-1)
\nonumber \\ && ~~~~
-f_{\rm init}(i+1,j) - f_{\rm init}(i-1,j)
]\bigr\}\int_{j-1/2}^{j+1/2} \phi\left( i-\tfrac12,y\right)\,dy
\nonumber \\ &&
+ \bigl\{(a_{\rm NN}+a_{\rm D})[f_{\rm init}(i,j+1)-f_{\rm init}(i,j)] + \tfrac12a_{\rm D}[f_{\rm init}(i+1,j+1) + f_{\rm init}(i-1,j+1)
\nonumber \\ && ~~~~
-f_{\rm init}(i,j+1) - f_{\rm init}(i,j-1)
]\bigr\}\int_{i-1/2}^{i+1/2} \phi\left( x,j+\tfrac12\right)\,dx
\nonumber \\ &&
+ \bigl\{ (a_{\rm NN}+a_{\rm D})[f_{\rm init}(i,j-1) - f_{\rm init}(i,j)]+ \tfrac12a_{\rm D}[f_{\rm init}(i+1,j-1) + f_{\rm init}(i-1,j-1)
\nonumber \\ && ~~~~
-f_{\rm init}(i,j+1) - f_{\rm init}(i,j-1)
] \bigr\}\int_{i-1/2}^{i+1/2} \phi\left( x,j-\tfrac12\right)\,dx
.
\label{eq:df}
\end{eqnarray}

A particular case of interest is a spot that is also ${\cal C}_{4v}$-symmetric and centered on a pixel -- say $(i,j)=(0,0)$ for simplicity. In this case, Eq.~(\ref{eq:df}) simplifies to
\begin{equation}
\frac{\Delta f(0,0)}{Q_{\rm tot}}
= 4\bigl[ a_{\rm NN}f_{\rm init}(1,0) + a_{\rm D}f_{\rm init}(1,1)  -(a_{\rm NN}+a_{\rm D})f_{\rm init}(0,0) \bigr]
\int_{-1/2}^{1/2} \phi\left(\tfrac12,y\right)\,dy.
\end{equation}
The $B$ coefficient defined in \citet{plazas18} (P2018) is this change of the flux in the central pixel divided by the contrast of the central pixel relative to the 4 nearest neighbors:
\begin{align}
B &= \frac{\Delta f(0,0)}{f(0,0)Q_{\rm tot}[f_{\rm init}(0,0) - f_{\rm init}(1,0)]} 
\nonumber \\
&= 4\frac{ a_{\rm NN}f_{\rm init}(1,0) + a_{\rm D}f_{\rm init}(1,1)  -(a_{\rm NN}+a_{\rm D})f_{\rm init}(0,0) }{ f_{\rm init}(0,0)[f_{\rm init}(0,0) - f_{\rm init}(1,0)] }
\int_{-1/2}^{1/2} \phi\left(\tfrac12,y\right)\,dy.
\label{eq:B}
\end{align}
Note that we would have $B = -4a_{\rm NN}$ in the limit of $a_{\rm D}\rightarrow 0$ (no diagonal BFE coupling) and $\phi(\tfrac12,y)\rightarrow f_{\rm init}(0,0)$ (uniform illumination). More generally, we may write
\begin{equation}
B = c_{\rm NN} a_{\rm NN} + c_{\rm D} a_{\rm D},
\end{equation}
where $c_{\rm NN}$ and $c_{\rm D}$ are coefficients that depend on the profile $\phi$.

\begin{figure}
\includegraphics[width=5.9in]{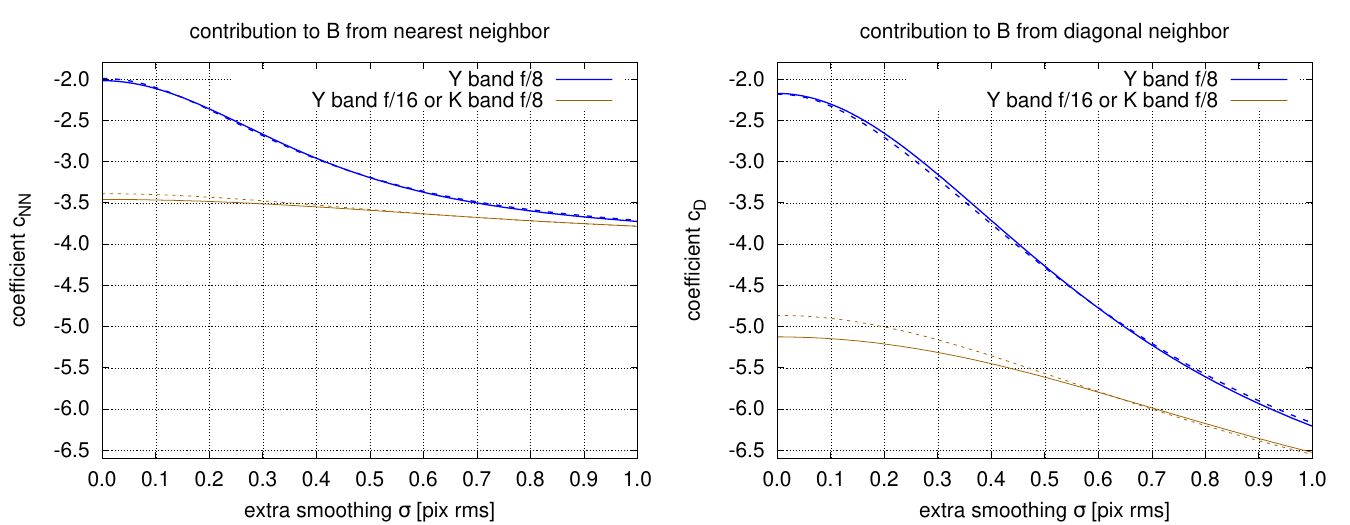}
\caption{\label{fig:c}The coefficients $c_{\rm NN}$ (left) and $c_{\rm D}$ (right), for two values of the diffraction spot size $\lambda f$. These are plotted as a function of the extra smoothing $\sigma$. The solid curves show the full numerical evaluation of the integral (Eq.~\ref{eq:B}), whereas the dashed curves show the Gaussian approximation (Eq.~\ref{eq:cGauss}) using a spot size that is the root-sum-square of the diffraction and Airy contributions, $\sigma_{\rm eff}^2 = \sigma^2 + (0.42\lambda f)^2$. Note that the uniform illumination limit ($c_{\rm NN}\rightarrow -4$ and $c_{\rm D}\rightarrow -8$) is not achieved for the range of spot sizes shown.}
\end{figure}

For a Gaussian spot with rms per axis $\sigma$ (before pixelization), we can analytically compute these coefficients:
\begin{equation}
c_{\rm NN} = -\frac{2\sqrt{2}\,e^{-1/(8\sigma^2)}}{\sqrt\pi\,\sigma\,{\rm erf}\, \frac1{2\sqrt2\,\sigma}}
~~~{\rm and}~~~
c_{\rm D} = -\frac{\sqrt{2}\,e^{-1/(8\sigma^2)}}{\sqrt\pi\,\sigma\,{\rm erf}\, \frac1{2\sqrt2\,\sigma}}
\left( 1 + 
\frac{
 {\rm erf}\, \frac3{2\sqrt2\,\sigma}
}{ {\rm erf}\, \frac1{2\sqrt2\,\sigma}}\right).
\label{eq:cGauss}
\end{equation}
However, for more general cases the coefficients must be computed numerically from Eq.~(\ref{eq:B}). We show in Fig.~\ref{fig:c} the values of these coefficients for an Airy disc convolved with a Gaussian (possibly representing charge diffusion or image motion). These are shown for a range of additional smoothing $\sigma$ (in rms pixels per axis), and for two values of the diffraction spot size $\lambda f$: the case considered in this experiment (Y band at f/8), and (for reference) a case twice as large (K band at f/8 or Y band at f/16).

\newpage

\bibliographystyle{mnras}
\bibliography{bfe_h4rg_ppl}

\end{document}